\documentstyle[mprocl]{article}

\input{epsf}
\def\Journal#1#2#3#4{{#1} {\bf #2}, #3 (#4)}

\def\AA{\em Astron. Astrophys.}
\def\APJ{\em Astrophys. J.} 
\def\JMP{\em J. Math. Phys.} 
\def\PRSLA{\em Proc. R. Soc. Lond. A} 
\def\zc{\bar{z}}

\begin{document}

\title{PROPERTIES OF POINT MASS LENSES ON A REGULAR POLYGON AND
THE PROBLEM OF MAXIMUM NUMBER OF IMAGES}

\author{ S. Mao}

\address{Max-Planck-Institut f\"ur Astrophysik,
Karl-Schwarzschild-Strasse 1,
\\ 85740 Garching, Germany}

\author{ A. O. Petters}

\address{Department of Mathematics, Princeton University, Princeton,\\
NJ, 08544-1000, USA}

\author{H. J. Witt}

\address{Astrophysikalisches Institut Potsdam, An der Sternwarte
16,\\14482 Potsdam, Germany}

\maketitle\abstracts{
We study the critical curves, caustics, and multiple
imaging due to one
of the simplest many-body gravitational lens
configurations: equal-mass point masses on the
vertices of a regular polygon.
Some examples of the critical curves
and caustics are also displayed. 
We pose the problem of determining the maximum number of
lensed images due to regular-polygon and general 
point mass configurations.  Our numerical simulations
suggest a maximum that is linear, rather than quadratic, 
in the number of point masses.
}
 
\section{Critical Curves and Caustics}

Suppose that 
$g$ point masses $m_k$, where
$m_k= 1/g$,  are on the vertices $z_k$ of a radius $r$ regular
polygon centered at the origin:
$z_k = r~e^{i {2 \pi (k-1) /g}},$
where $k = 1, \dots , k$ and $g\ge 2$.
[A point mass lens ($g=1$) produces a circular critical curve
and point caustic;  light sources off the point caustic have
two lensed images.]  
The associated {\it lens equation},
expressed in complex quantities (Bourassa, Kantowski \&
Norton,~\cite{BKN}
Witt~\cite{Wtt90}),
is given by
\begin{equation}\label{lensEq}
z_s =  z - \sum_{k=1}^g {m_k \over \zc - \bar{z_k}}
= z - {\zc^{g-1} \over \zc^g - r^g},
\end{equation}
where $z_s$ is the light-source position. The lens equation
defines a mapping, $\eta: z\mapsto z_s$, 
from the lens plane into the light source plane.  {\it Lensed images}
of a light source at $z_s$ are solutions $z$ in ${\bf C}$ of the
lens equation.
{\it Critical curves} (i.e., set of all infinitely magnified lensed
images) are given by setting the Jacobian determinant $J$ of
$\eta$ equal to zero:
$$J = 1 - {\partial z_s \over \partial \zc} \overline{\partial z_s \over
\partial \zc} = 0.$$
This is solved by $\frac{\partial z_s}{\partial \zc} = e^{i \phi}$,
where $0 \le \phi < 2 \pi.$  
The critical curves are then the solution curves
$z(\phi)$, where $0 \le \phi < 2 \pi$, of
\begin{equation}
\label{ccpoly}
p(z) = z^{2g} + e^{i \phi} z^{2g-2}  -2 r^g z^g + 
(g-1) r^g e^{i \phi} z^{g-2} + r^{2g}
= 0.
\end{equation}
Since $z=0$ is not a root of Eq.(\ref{ccpoly}),
{\it no critical curve and no caustic passes through the origin.}
{\it Caustics} (i.e., set of positions
from which a light source has at least one infinitely magnified lens
image) are the $\eta$-images of critical curves.
By Eq.(\ref{ccpoly}), {\it there are at most $2g$ critical curves; hence,
the same for caustics.}
If critical curves merge for a given parameter value,  
then $p(z)$ has a double  or higher order
zero.  But ${\partial^2 z_s \over \partial \zc^2}$  is equivalent to
a complex polynomial in $z$ of degrees $3$ and $6$ for  $g=2$ and $3$,resp.,
and degree
$2g+1$ for $g\ge 4$. In the latter case, one solution is $z=0$, which
cannot lie on no critical curve or caustic.   
Thus, {\it if $g=2,3$, then there are  at most
$3, 6$ beak-to-beak caustics, resp., while at most $2g$ occur for 
$g\ge 4$.}  A general $g$-point mass system has at most
$3g-3$ beak-to-beaks (Witt \& Petters~\cite{WP}).

Examples of the critical curves and 
caustics are shown in Figure 1 as a function of $r$ 
for the case $g=6$.

\begin{figure}[tbh]
\epsfxsize=3.3in\hspace{.75in}\epsfbox{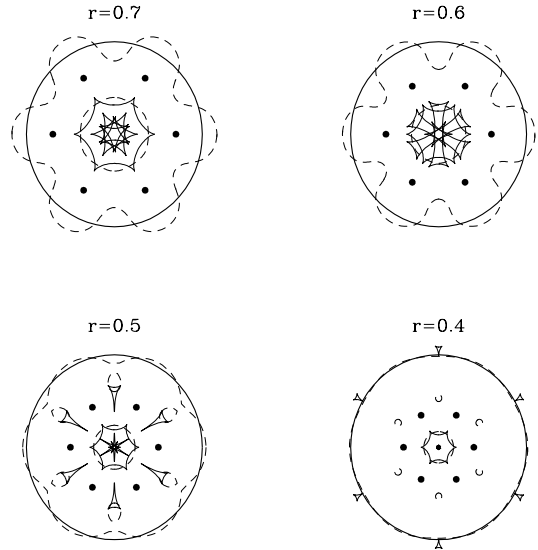}
\caption{ Critical
curves (dashed lines) and caustics (thin solid lines)
are shown for equal-mass sextuple lenses (filled dots) 
as a function of the radius of the regular polygon.
A unit circle
is indicated with the thick solid line.
}
\label{polymg8fig1}
\end{figure}

\section{Number of Lensed Images of a Light Source at the Center}

For a light source at the center of the polygon ($z_s=0$), the lens equation
becomes
\begin{equation} \label{lensEq2}
\rho = { \rho^{g-1} \over \rho^g - r^g e^{i g \theta}},
\end{equation}
where we expressed $z$ in the polar coordinates $\rho e^{i\theta}$.
An immediate solution is $\rho=0$. For any other solution, $e^{i g\theta}$
must be real and, hence, is either $+1$ or $-1$.  Eq.(\ref{lensEq2})
can be recast into $f_{\pm} (\rho) \equiv \rho^g-\rho^{g-2} \pm r^g =0$,
where $f_+$ and $f_-$ correspond to the lens equation
with $e^{i \theta}=-1$ and $+1$, resp.
If $n_\pm$ are the number of positive zeros of
$f_\pm$, then the total number of images is then
simply given by, $N= g(n_+ + n_-)+1$, where the factor of $g$ arises
due to rotational symmetry.  
If $g=2$, then
$n_+ =  0,1$
for  $r\ge1$ and
$r<1$, resp., and
$n_- = 1$.  {\it It follows that  $N = 3$ or $5$ if $g=2$.} 
Now, suppose that $g\ge 3$. 
By Descartes' rule of signs, we have
$n_\pm \le \ \# (\mbox{sign changes in coefficients of}\ f_\pm)$.  
Consequently,  $n_+\le 2$ and $n_-\le 1$ for $g\ge 3$.  
Hence, $N\le 3g +1$.  This upper bound is also the maximum, that is,
it is
achievable for each $g$.  In fact, 
let $r_{cr} = \left(\rho_m^{g-2}-\rho_m^g\right)^{1/g}$, where
$\rho_m=[(g-2) / g]^{1/2}$.
It can be shown that if $g\ge 3$,
then $n_- =1$ and 
$n_+ = 0, 1, 2$ for \ $r> r_{cr}$, 
$r= r_{cr}$,
$r< r_{cr}$, resp.
Thus,  {\it if $g\ge 3$, then 
$N =g+1, 2g +1, 3g+1$ 
for 
$r> r_{cr}$,
$r= r_{cr}$, and
$r< r_{cr}$, resp.}

\section{Open Problem}

The bounds on the total number of lensed
images due to $g$ point masses (not necessarily  on a regular polygon)
are known to be
$g +1 \le N \le g^2+1$.
The lower bound follows rigorously from Morse theory (Petters~\cite{Ptt92}),
while the upper bounds can be shown using a trick substitution
(Witt --- see Ref. 2), or, via 
resultants (Petters~\cite{Ptt97}).  
The lower bound $g+1$ is achievable for each 
$g$ (Petters~\cite{Ptt96}); hence, it is the 
minimum number of lensed images.  We do not know whether the upper
bound $g^2 +1$ is attainable for each $g$.
For $g$ point masses on the vertices of a regular polygon, and
light source not necessarily at the origin, 
the maximum number
of images appears to be $3g+1$. Our numerical simulations for generic
point-mass configurations seem to confirm this limit as well. 
{\it It is unknown to the authors whether 
the maximum number of lensed images due to a general $g$ point mass
lens system  
is linear, i.e., does 
$N_{max} = g n_1  + n_2$ for each $g\ge 1$, 
where $n_1$ and $n_2$ are fixed positive
integers?}

\section*{Acknowledgments}
S.M. is partly supported by the ``Sonderforschungsbereich
375-95 f\"ur
Astro--Teil\-chen\-phy\-sik" der Deutschen For\-schungs\-ge\-mein\-schaft.
A.P. was supported in part by NSF Grant No. DMS-9404522.

\end{document}